\documentclass[]{article}
\usepackage{amsmath}
\usepackage{amsfonts}
\usepackage{amssymb}
\usepackage{graphicx}
\usepackage{amsfonts}
\usepackage[english]{babel}
\usepackage[utf8]{inputenc}
\usepackage{times}
\usepackage[T1]{fontenc}

\begin{document}

\title{Lorentz Spin-Foam with Non Unitary Representations by use of Holomorphic Peter-Weyl Theorem}
\author{Leonid Perlov\\
Department of Physics, University of Massachusetts,  Boston\\
leonid.perlov@umb.edu
}
\date{ October 4, 2015}

\maketitle

\begin{abstract}
In quantum gravity the unitary evolution does not follow from the Wheeler-DeWitt dynamics equation as it follows from the Schrödinger equation in non-relativistic quantum mechanics. Therefore we can define a spin-foam model based on SL(2,C) spinor finite non-unitary representations. The recently discovered holomorphic Peter-Weyl theorem \cite{Huebschmann} made it possible to decompose  the delta function of a non-compact Lorentz group into the convergent sum of the matrix coefficients. We calculate the vertex amplitude with the help of that theorem and obtain a simple expression for our model. The $SL(2,C)$ Hilbert space is defined from $SU(2)$ Hilbert space by  Huebschmann-Kirillov transform \cite{Huebschmann}. A new transform is simpler than the well known Hall transform as it does not contain a heat kernel convolution.  We do not set Barbero-Immirzi constant $\gamma$ a priori,  instead we obtain it as a solution of the diagonal and off-diagonal simplicity constraints being $\gamma = \frac{-in}{(|n| + 2p)}$ where $p$ is a non-negative half-integer. When $p=0$ the solution corresponds to the Ashtekar's self-dual connections. We point out that the Barbero-Immirzi becomes real when one chooses a unitary representation. It is complex when the representation is non-unitary principal series or non-unitary spinor representation. 
\end{abstract}

\section{Introduction}
The Lorentz spin-foam models were considered in several papers \cite{Pereira2008} \cite{EPRL} \cite{BarrettCrainLorentz} \cite{Bianchi}. All previous models were based on the unitary principal series, i.e the unitary infinite dimensional Lorentz group representation \cite{Gelfand} \cite{Naimark} \cite{Ruhl}. In \cite{Pereira2008} the vertex amplitude expression contains the integral with the Plancherel measure. The expression has a simple form however it is divergent and, as mentioned by the authors in the discussion, is formal. Another unitary Lorenzian model \cite{BarrettCrainLorentz} is convergent with some additional regularization procedure. Further progress in the holomorphic approach has been made in \cite{Bianchi} by use of the Hall transform. The unitary choice of the representations in all above models implies that the principal series representation  parameter $\rho$ is real rather than complex as it would be in the case of non-unitary principal series. As a consequence the Barbero-Immirzi constant becomes real, which can be seen directly from the simplicity constraint solution $\rho = \gamma n$, where $\gamma$ -  Barbero-Immirzi constant, $n$ - is a positive integer. We call it an implicit reality condition on the Barbero-Immirzi constant. In quantum gravity the unitary evolution does not follow from the dynamics equation as it does in non-relativistic quantum mechanics. One can see it directly from the Wheeler-DeWitt and Schrödinger equations. Therefore in the current paper we use non-unitary finite dimensional $SL(2,C)$ representations to define a spin-foam model. We solve the simplicity constraints with respect to the Barbero-Immirzi parameter and obtain the series of solutions  $\gamma = \frac{-in}{(|n| + 2p)}\;$,  $j_-=p/2$, $j_+=j+ p/2$,  where $p$ - is a non-negative half-integer, $j_-$ and $j_+$ - are $SL(2,C)$ spinor representation parameters,  $j$ is an $SU(2)$ spin. When $p=0$ the solution corresponds to the model with the self-dual connection. In addition the Huebschmann-Kirillov transform and the the holomorphic Peter-Weyl theorem  \cite{Huebschmann} let us find the inner product and the corresponding orthonormal basis for the Lorentz spinor representation Hilbert space. The spin-foam vertex amplitude is then calculated by decomposing the delta function into the converging sum of  $SL(2,C)$ matrix coefficients. As a result we obtain a simple vertex amplitude expression and a series of solutions with respect to   Barbero-Immirzi cosntant $\gamma$ and $SL(2,C)$ spinor representation spins $j_-$ and $j_+$.\\[2ex]
	The paper is organized as follows. In \ref{subsec:NonunitarySpinfoam} we define the spin-network and spin-foam by using $SL(2,C)$ non-unitary spinor representations. We then use the holomorphic Peter-Weyl theorem  to define the Hilbert space spanned by $SL(2,C)$  matrix coefficients.  In \ref{subec:SimplicityConstraints}  we solve the simplicity constraints by calculating the $SL(2,C)$ spinor representation Casimir and pseudo-Casimir invariants.  In  \ref{subsec:VertexAmplitude} we proceed by calculating the spin-foam vertex amplitude. The last chapter \ref{sec:Discussion} is a discussion. Appendix A contains  the details on the spinor non-unitary finite representations of the non-compact groups and their maximal compact subgroups. Appendix B contains Casimir and pseudo-Casimir derivation for $SL(2,C)$ spinor representations by use of the differential operators. 

\section[test]{Non-unitary Lorentz Spin-Foam}
\label{sec:NonunitarySpinfoam}

\subsection{Non-Unitary Lorentz Spin-Networks and Holomorphic Peter-Weyl Theorem}
\label{subsec:NonunitarySpinfoam}

Consider a four dimensional Lorentz analytic manifold M.  We define a spinor non-unitary spin-network in a conventional way as a one dimensional oriented  graph $\varGamma$, where each link  is assigned a $2j+1$ dimensional non-unitary  regular $SL(2,C)$ representation and  each vertex is assigned an intertwiner between the tensor products of the representations assigned to the incoming and outgoing edges.
The $SL(2,C)$ spinor representation matrix coefficients $D^j_{q, q'}(g), g \in SL(2, C)$ are  the analytic continuation of the $SU(2)$ spinor matrix coefficients $D^j_{qq'}(u), u \in SU(2)$. The complete basis is spanned by the tensor product of the two matrix coefficients  $D^{(j_- j_+)}_{q_- q'_- q_+ q'_+} :=D^{j_-}_{q_-, q'_-}(\bar g) \otimes D^{j_+}_{q_+, q'_+}(g), g \in SL(2, C)$, $\bar g$ - is a complex conjugate.  
The spin-network function is then defined as contraction of the spinor matrix-coefficients $D^{(j_- j_+)}_{q_- q'_- q_+ q'_+}(g)$ with the intertwiners $i_n$:
\begin{equation}
\label{eq:cylindricalfunctiontensor}
\varPsi(A)= \bigotimes\limits_{k=1}^{L} D^{(j_- j_+)}_{q_- q'_- q_+ q'_+}(h(A, \gamma_k)) \bigotimes\limits_{n=1}^{N} i_n
\end{equation}
,where $A$ is a connection defined in each point of the spin-network, $\gamma_k$ is a spin-network edge, $h(A, \gamma_k)$ is a holonomy, i.e parallel transport along $\gamma_k$. \\

In order to define the inner product on the Hilbert space of such functions, we are going to use the holomorphic Peter-Weyl theorem \cite{Huebschmann}, which establishes the isomorphism between the Hilbert space spanned by the compact group K representation matrix coefficients and the Hilbert space spanned by the matrix coefficients of that group complexification $K^{\mathbb{C}}$:
\begin{equation}
\label{matrixcoefficientmap}
\phi^{\mathbb{C}}(g)  \rightarrow  {(\hbar\pi)}^{{dim(K)}/4}e^{\hbar{|\lambda + \rho|}^2/2}\phi(g)
\end{equation}
We call it Huebschmann-Kirillov transform. The inner products of these two Hilbert spaces are related as follows:
\begin{equation}
\label{eq:innerprodgeneral}
\int\limits_{K^{\mathbb{C}}}\bar\phi^{\mathbb{C}}(g)\phi^{\mathbb{C}}(g) e^{-|Y|^2/\hbar} \eta(g) du \,dY = {(\hbar\pi)}^{{dim(K)}/2}e^{\hbar{|\lambda + \rho|}^2} \int\limits_K \bar\phi(u)\phi(u)\, du
\end{equation}
,where $g \in K^{\mathbb{C}}$ and we use polar decomposition $g = ue^{iY}, u \in K, Y \in t, $ algebra of K,\\
$\lambda$ is the highest weight of $K$, while $\rho$ is the Weyl vector of $K$, i.e the half sum of the positive roots, 
the density of the measure on the left hand side is
\begin{equation}
\label{eq:measure}
\eta(u, Y) = \left(det\left(\frac{sin(ad(Y))}{ad(Y)}\right)\right)^{\frac{1}{2}},  u \in K,  Y \in t
\end{equation}

By using (\ref{eq:measure}) we can calculate the inner product of the non-compact group representation Hilbert space by calculating the inner product of its isomorphic projection to the Hilbert space of its maximum compact subgroup representation. Since the map is provided by constant multiplication, all the orthonormal properties of the matrix coefficients in the compact group case propagate to the Hilbert space of its non-compact complexification. This map is  much simpler than the well known Hall transform as it does not contain the heat kernel convolution. Even though the inner product is not Lorentz invariant, it is $SU(2)$ invariant and provides a map between the $SU(2)$ invariant physical space obtained as the solution of the simplicity constraint and its image in the Lorentz space.

In our case the compact group $K$ is $SU(2)$ and its complexification $K^{\mathbb{C}}$ is $SL(2,C)$.  The simplicity constraints solution  below provides the following matrix coefficients map:


\begin{equation}
\label{eq:representativefunction}
\phi(u) = D^j_{qq'}(u) , \qquad  \phi^{\mathbb{C}}(g) =  D^{(p/2, j + p/2)}_{qq'}(g) 
\end{equation}\\

where $u \in SU(2), g \in SL(2,C)$, $j$ - is a non-negative half integer $SU(2)$ spin, $ p = 0, 1, \mbox{...} $. 
The above map of $SU(2)$ spin $j$ to $SL(2,C)$ spins $\left(j_- = p/2, j_+= j + p/2\right)$  is obtained below in the section  ($\ref{subec:SimplicityConstraints}$) as the simplicity constraints solution. \\


The weight constant in ($\ref{matrixcoefficientmap}$)
\begin{equation}
C_j = {(\hbar\pi)}^{{dim(K)}/4}e^{\hbar{|\lambda + \rho|}^2/2}
\end{equation}

 in our case of $K=SU(2), K^{\mathbb{C}}= SL(2,C)$ is calculated in the following manner. The $\dim(K) = 3$, the highest weight $\lambda$ of the finite dimensional representation is $( \dim(V) - 1) \frac{\alpha(H)}{2}$, $V$ is the representation vector space. Since $dim(V) = 2j + 1$,  it follows that $\lambda = 2j \frac{\alpha(H)}{2}$, where $\alpha(H)$ is the only $SU(2)$ positive root $\alpha(H) = 2h$, $H = \mbox{diag}(ih, -ih)$. The Weyl vector $\rho = \frac{\alpha(H)}{2}$. The Killing form gives the value of ${|\lambda + \rho|}^2 =\frac{{(2j + 1)}^2}{8}$. By substituting these values into ($\ref{matrixcoefficientmap}$) we find the multiplier constant:
\begin{equation}
\label{holomorphicoeff}
C_j = {(\hbar\pi)}^{3/4}e^{\frac{\hbar{(2j + 1)}^2}{8}}
\end{equation}\\

The inner products relation  ($\ref{eq:innerprodgeneral}$) for our case of $K=SU(2), K^{\mathbb{C}}= SL(2,C)$  becomes

\begin{equation}
\label{eq:innerprod}
\int\limits_{K^{\mathbb{C}}}\overline{D^{\left(p/2, \; j + p/2 \right)}_{qq'}(g)}D^{\left(p/2, \; j + p/2 \right)}_{qq'}(g) e^{-|Y|^2/\hbar} \eta(g) du \,dY =  {C_j}^2 \int\limits_K \overline{D^j_{qq'}(u)}D^j_{qq'}(u)\, du
\end{equation}\\

Since the Wigner matrices $D^j_{qq'}(u), u \in SU(2)$  provide the orthonormal basis, the same property by (\ref{eq:innerprod}) propagates  to $K^{\mathbb{C}}$ matrix coefficients $D^{(p/2, j + p/2)}_{qq'}(g)$ . It follows then from (\ref{eq:cylindricalfunctiontensor}) that the same is true for the spin-network (cylindrical) functions defined on the spinor representations of the Lorentz group $\varPsi(A)$.

By the holomorphic Peter-Weyl theorem \cite{Huebschmann}, the matrix coefficients $\phi^{\mathbb{C}}(g)$ are dense in $L^2(g, \eta(g) du \,dY )$. It implies that the spin-network functions $\varPsi(A)$ from  (\ref{eq:cylindricalfunctiontensor}) are also dense in this space.
So we receive the orthonormal basis of the  spin-network functions of the Hilbert space $H_\varGamma$.

\begin{equation}
\varPsi(A)=\prod\limits_{k=1}^{L}{ D^{(p/2, j + p/2)}_{q_k q'_k}(g)  (h(A, \gamma_k)) }\prod\limits_{n=1}^{N} {i_n}
\end{equation}
,where $i_n$ - the intertwiners between the tensor product of the representations assigned to the vertex incoming and outgoing edges. The intertwiners contract with the representations  as in a standard spin-network definition.\\[2ex]

\subsection{Spinor Lorentz Spin-Foam and Simplicity Constraints}
\label{subec:SimplicityConstraints}

We consider the spin-foam defined on the spin-networks discussed in the previous section. The simplicity constraints  are in the form introduced  in EPRL model \cite{EPRL}, i.e expressed through the representation Casimir  $C_1$ and pseudo-Casimir $C_2$. 
The diagonal and off-diagonal constraints are: 
\begin{equation}
\label{diagonal}
C_2\left(1-\frac{1}{\gamma^2}\right) + \frac{2}{\gamma}C_1 \approx 0
\end{equation}
\begin{equation}
\label{offdiagonal}
C_2=4\gamma L^2
\end{equation}
,where $C_1 = J \cdot J$ - Casimir Scalar    and $C_2 = \,^*J \cdot J$ - Casimir pseudo-scalar, L - rotation generators.


 $SL(2,C)$ spinor representation Casimir and pseudo-Casimir are derived by using the differential operators in Appendix B, where we arrive at the following expressions:

\begin{equation}
\label{casimirs}
C_1 = 4(j_+( j_+ +1) + j_-( j_-+1))  \quad  C_2 =-4i( j_+(j_++1) - j_-(j_-+1) )  
\end{equation}

After selecting the spectrum as in \cite{EPRL} , i.e  $j_+^2$  instead of $j_+(j_++1)$,  and substituting the Casimir and pseudo-Casimir from  (\ref{casimirs}) into  (\ref{diagonal})  and  (\ref{offdiagonal}) the  constraints become:

\begin{equation}
\label{diagonal1}
-4i (j_+^2 -j_-^2) ( \gamma - \frac{1}{\gamma}) = -2 (4 (j_+^2 + j_-^2))
\end{equation}

\begin{equation}
\label{offdiagonal1}
-4i (j_+^2 - j_-^2) = 4 \gamma j^2
\end{equation}

From (\ref{offdiagonal1}) we can immediately see that $\gamma$ is pure imaginary. Let $\gamma = m i, \; m \in R $
Then  (\ref{offdiagonal1})  becomes

\begin{equation}
\label{offdiagonal2}
j^2_-= m j^2 + j^2_+
\end{equation}
When solving with respect to $m$, we obtain:
\begin{equation}
\label{em}
m = \frac{j^2_- - j^2_+}{j^2}
\end{equation}
Since the spinor representation is part of the principal series non-unitary representation with the parameters $(n, \rho), \; n \in Z, \rho \in C$ one can express the half integer spins $(j_-, j_+)$ via $(n, \rho)$ by using the following relations $\cite{Naimark}$ p. 295:
\begin{equation}
\label{spins}
2j+ = \frac{|n|}{2} + \frac{i \rho}{2} \quad
2j_- =  -\frac{|n|}{2} + \frac{i \rho}{2}
\end{equation}
where $\rho$ now for the finite dimensional representations is:
\begin{equation}
\label{ChangingOrder}
\rho = -i (|n| + 2p), \;\; p = 0, 1, 2 \mbox{...}
\end{equation}
it follows that:
\begin{equation}
\label{nro}
n = (2j_+ - 2j_-),   \quad
i\rho = ( 2j_+ + 2j_-)
\end{equation}
By substituting  ($\ref{ChangingOrder}$) into  ($\ref{spins}$) we obtain:
\begin{equation}
j_- = p/2  \quad j_+ = |n|/2 + p/2    
\end{equation}
By substituting  ($\ref{ChangingOrder}$) into the first diagonal simplicity constraint solutions $\rho = n \gamma $ or $ \gamma = \rho / n $ we obtain:
\begin{equation}
\label{gamma6}
\gamma = \frac{-i(|n| + 2p)}{n}
\end{equation}
However the off-diagonal constraint: $n \rho = 4 \gamma L^2$ produces the solution $n = 2(|n|/2 + p)$ it follows that $p = 0$ and the only solution for $\gamma$ in this case as it follows from ($\ref{gamma6}$) is $\gamma = \pm i$.
By substituting  ($\ref{ChangingOrder}$) into the second diagonal simplicity constraint solutions $\gamma = \frac{-n}{\rho}$ we obtain:
\begin{equation}
\gamma = \frac{-in}{(|n| + 2p)}
\end{equation}
This solution for $\gamma$ also contains $\gamma = \pm i$, when $p = 0$. Therefore it contains all solutions for $\gamma$ corresponding to the spinor representations. 
We can now express our newly found solution of the simplicity constraints via spins
\begin{equation}
( p/2, \; |n|/2 + p/2), \; n \in Z, p = 0, 1, 2, \; \mbox{ ... }
\end{equation}
By using the $SU(2)$ half integer spin $j = \frac{|n|}{2}$, corresponding to the lowest representation we can write the solution as:
\begin{equation}
\label{solutioninj}
(p/2, \; j + p/2) \; , \;  \quad  \gamma = -\mbox{sgn}(n) \, i \, \frac{j }{j  + p }
\end{equation}

Solving (\ref{diagonal1}) and (\ref{offdiagonal2}) with respect to $j_-, j_+$ and $m$ we obtain the following solution:

\begin{equation}
\label{j-}
j_-= p/2
\end{equation}

\begin{equation}
\label{j+}
j_+ = j + p/2
\end{equation}
Also we can see that $m$ in ($\ref{em}$)  $\gamma = mi$ equals to $m = -\mbox{sgn}(n) \, \frac{j }{j  + p }$


We would like to point out the special case of $p=0$. The solution in that case is  $\gamma = \pm i, \; (0, j)$, which corresponds to the  Ashtekar's self-dual connection.


\subsection{Spinor Vertex Amplitude and Transition Matrix}
\label{subsec:VertexAmplitude} 

We applied the holomorphic Peter-Weyl theorem $\cite{Huebschmann}$ to map $SU(2)$ matrix coefficients to the physical Hilbert space provided by the simplicity constraint solution.



\begin{equation}
D^j_{qq'}(u)  \rightarrow \frac{1}{C_j} D^{(p/2, \; j + p/2)}_{qq'}(g) 
\end{equation}

or by substituting $C_j$ from ($\ref{holomorphicoeff}$)

\begin{equation}
D^j_{qq'}(u)  \rightarrow  {(\hbar\pi)}^{-3/4}e^{- \frac{\hbar{(2j + 1)}^2}{8}} D^{(p/2, j + p/2)}_{qq'qq'}(g)  
\end{equation}\\

The amplitude of the vertex bound by ten faces, each assigned $SU(2)$ spin $j_{ab}$ and the edges $SU(2)$ spins $i_a$ is then similar to EPRL Euclidean $SO(4)$ case $\cite{EPRL}$. However 15j symbol spins are different and depend on a half-integer m. 



\begin{equation} 
A = \sum\limits_{i_+^a i_-^a} 15j \left ( p/2,  i_-^a \right) 15j \left ( j_{ab} + p/2,  i_+^a\right) 
\end{equation}

\section{Discussion}
\label{sec:Discussion}

We have defined the Lorentzian spin-foam model by using the non-unitary finite dimensional  $SL(2,C)$ spinor representations. The simplicity constraints in this case produced the solution $\gamma = \frac{-in}{(|n| + 2p)} \;$,   $j_-= p/2 \;$, $j_+ = j + p/2\;$,   $j$ - $SU(2)$ spin. This solution corresponds to the self-dual connection, when $p=0$. We have defined the Hilbert space spanned by $SL(2,C)$ spinor representation matrix coefficients and introduced the inner product by using the recently discovered holomorphic Peter-Weyl theorem $\cite{Huebschmann}$.The same theorem provides the Huebschmann-Kirillov transform between the Hilbert space spanned by the compact group matrix coefficients and the Hilbert space of the compact group complexification. The new transform has a much simpler form than the previously known Hall-Bargman-Segal transform. The latter uses the function convolution with the heat kernel \cite{Hall} \cite{Ashtekar-Hall}, while in the Huebschmann-Kirillov transform the map is provided by weight multiplication. The $SL(2,C)$ Hilbert space inner product is then defined via the corresponding $SU(2)$ inner product multiplied by the weight ${C_j}^2$ which depends only on the $SU(2)$ spin $j$.  Even though the inner product is not Lorentz invariant, it is bi-$SU(2)$ invariant and can be used on the physical space that is a projection of the Lorentz space to its $SU(2)$ subspace provided by the solution of the simplicity constraints. As a result the vertex amplitude expression has a much simpler form as it does not contain the Clebsch-Gordan coefficients. Also the $SL(2,C)$ delta function decomposition into $SL(2,C)$ matrix coefficients, used in the vertex amplitude derivation is now well defined and convergent.

\section{Appendix A - $SL(2,C)$ Spinor Finite Non-Unitary versus Infinite Unitary Lorentz Representations}
\label{sec:AppendixA}
	As it is well known the Lorentz group is non-compact. Therefore the only unitary representations are  the principal and complementary series $T^{n\rho}(g)$ on the vector space of the homogeneous functions. These representations are infinite dimensional. 

In the present model we use the finite dimensional  Lorentz non-unitary regular right representations which are similar to the regular right $SU(2)$ representation on the vector space of the homogeneous polynomials with the complex coefficients. \\
Let $\xi$ be a complex two vector $\xi = (z_1, z_2)$  that transforms as ${\xi}'=\xi g, \; g \in SL(2,C)$. Consider the vector space of homogeneous polynomials $p(z_1, z_2)$ of degree $2j$ in $z_1$ and $z_2$, and $j$ being a half-integer. In this space of the dimension $(2j+1)$ we define a spinor transformation $D(g$) for any $g \in SL(2,C)$ \\

$D(g)p(\xi) = p(\xi g)$\\

This transformation provides  $(2j+1)$ dimensional representation of the group 

$ SL(2, C)$:
$
\begin{pmatrix}
a  & b \\
c & d\\
\end{pmatrix}
$
and $ad - bc=1$\\[2ex]

$SU(2)$ is the maximal compact group of $SL(2,C)$. Its spinor representation is exactly the same as $SL(2,C)$ with $a,b,c$ and $d$ replaced by the components of 
\\[2ex]
$SU(2)=$
$
\begin{pmatrix}
\alpha + i\beta & \gamma + i \delta \\
-\gamma + i\delta &  \alpha - i\beta\\
\end{pmatrix}
$\\[2ex]
,where $ \alpha, \beta, \gamma$  and $ \delta $ are real, and  $\alpha^2 + \beta^2 + \gamma^2 + \delta^2 =1$\\

Replacing $ \alpha, \beta, \gamma$  and $ \delta $ with the complex $g_{ij}$ is a complexification of $SU(2)$ group.\\[2ex]
The matrix coefficients of the spinor $(2j+1)$ dimensional representations of $SU(2)$ and SL(2,C ) have the same form:

\begin{multline}
D^j_{qq'}(g) =  \left[\frac{(j+q)!(j-q)!}{(j+q')(j-q')!}\right]^{1/2} \sum\limits_n \begin{pmatrix} j+q' \\ n \end{pmatrix} \begin{pmatrix} j-q' \\j+q-n\end{pmatrix}
\\\times g_{11}^n g_{12}^{j+q-n}g_{21}^{j+q'-n}g_{22}^{n-q-q'}
\\\mbox{ ,where }  q\leq |j| 
\end{multline}

The complete basis is spanned by the tensor product of the two matrix coefficients:
\begin{equation}
D^{(j^- j^+)}_{q^+ q'^+ q^- q'^-} :=   D^{j^-}_{q^-, q'^-}(\bar g) \otimes D^{j^+}_{q^+, q'^+}(g), g \in SL(2, C)
\end{equation}

We receive $SU(2)$ matrix coefficients $D^j_{qq'}(u)$ for $ u \in SU(2) $ by restricting $g$ to $u$. 
Any $SL(2,C)$ finite dimensional irreducible representation is equivalent to the spinor representation. For further details refer to \cite{Ruhl} \cite{Carmeli}

\section{Appendix B  - Casimir and pseudo-Casimir for $SL(2,C)$ Spinor Non-Unitary Representations}

We use the  $SL(2,C)$ spinor representation differential operators to calculate the corresponding Casimir and pseudo-Casimir.  (See \cite{Ruhl} \cite{Carmeli})
\begin{equation}
\label{CasimirSpinor}
C_1 = L_+L_- + L_-L_+ + 2L_3^2 - K_+K_- - K_-K_+ - 2K_3^2
\end{equation}

\begin{equation}
\label{pseudoCasimirSpinor}
C_2= - (L_+K_- + L_-K_+ +  K_+L_- + K_-L_+ + 4L_3K_3)
\end{equation}

We use the following $SL(2,C)$ non-unitary representation differential operators:

\begin{equation}
\label{FirstDifOperator}
L_+ \, p= \left[ -\frac{\partial}{\partial z} -{ \bar z}^2 \frac{\partial}{\partial {\bar z}}  + n \bar z \right] \, p
\end{equation}

\begin{equation}
L_- \, p = \left[ z^2 \frac{\partial}{\partial z } + \frac{\partial}{\partial \bar z} -mz \right] \, p
\end{equation}

\begin{equation}
L_3 \, p = \left[-z\frac{\partial}{\partial z } + \bar z \frac{\partial}{\partial \bar z} + \frac{1}{2}(m-n) \right] \, p
\end{equation}

\begin{equation}
K_+ \, p = \left[i\frac{\partial}{\partial z} - i {\bar z}^2\frac{\partial}{\partial \bar z} + in \bar z \right] \, p
\end{equation}

\begin{equation}
K_- \, p = \left[-iz^2\frac{\partial}{\partial z } + i \frac{\partial}{\partial \bar z} + imz \right] \, p
\end{equation}

\begin{equation}
\label{LastDifOperator}
K_3 \, p = \left[iz\frac{\partial}{\partial z} + i \bar z \frac{\partial}{\partial \bar z} - \frac{i}{2}(m+n)\right] \, p
\end{equation}

By substituting (\ref{FirstDifOperator}) - (\ref{LastDifOperator}) into the expressions for Casimir  (\ref{CasimirSpinor})  and pseudo-Casimir  (\ref{pseudoCasimirSpinor}), doing the calculations, and at the end using the half integers $j_+$ and $j_-$ instead of $m$ and $n$
\begin{center}
$ n=2j_+ \qquad m=2j_-$
\end{center}

we obtain:

\begin{equation}
C_1 = 4(j_+( j_++1) + j_-( j_-+1))  
\end{equation}

\begin{equation}
C_2 =-4i( j_+(j_++1) - j_-(j_-+1) ) 
\end{equation}

\begin{center}
\line(1,0){50}
\end{center}

It's a pleasure to thank  Prof. Olchanyi,  Michael Bukatin, Leon Peshkin and Virginia Savova for multiple discussions.

\end{document}